\documentclass{llncs}
\usepackage{makeidx}  
\usepackage{amssymb}
\usepackage{url}
\usepackage{graphicx}
\usepackage{color}                           
\usepackage{epstopdf}

\def\R{\mathbb R}

\begin{document}
\mainmatter

\title{Reduction of the Pareto Set in Bicriteria Asymmetric Traveling Salesman Problem}
\author{Aleksey O. Zakharov\inst{1}\orcidID{0000-0003-2469-023X} \and \\
Yulia V. Kovalenko\inst{2,3}\orcidID{0000-0003-4791-7011}}

\authorrunning{A. Zakharov \and Yu. Kovalenko} 
\titlerunning{Reduction of the Pareto Set in bi-ATSP}
\institute{Saint Petersburg State University, St. Petersburg, Russian Federation \\
\email{a.zakharov@spbu.ru}
 \and
 Department of Mechanics and Mathematics, Novosibirsk State University, Novosibirsk, Russian Federation
 \and
  Sobolev Institute of Mathematics, Novosibirsk, Russian Federation \\
\email{julia.kovalenko.ya@yandex.ru}
}
\maketitle

\begin{abstract}
We consider the bicriteria asymmetric traveling salesman problem (bi-ATSP).
Optimal solution to a multicriteria problem is usually supposed to be the Pareto set,
which is rather wide in real-world problems.
We apply to the bi-ATSP  the axiomatic approach of the Pareto set reduction proposed by V.~Noghin.
We identify series of ``quanta of information'' that guarantee the reduction of the Pareto set for particular cases of the bi-ATSP.
An approximation of the Pareto set to the bi-ATSP is constructed by a new multi-objective genetic algorithm.
The experimental evaluation carried out in this paper
shows the degree of reduction of the Pareto set approximation for various ``quanta of information''
and various structures of the bi-ATSP instances generated randomly.

\end{abstract}

\section{Introduction} \label{sec:intro}
The asymmetric traveling salesman problem (ATSP) is one of the most popular problems in combinatorial optimization~\cite{ACGKMP2005}.
Given a complete directed graph where each arc is associated
with a positive weight, we search for a circuit visiting every vertex of the graph exactly once and minimizing the total weight.
In this paper, we consider the bicriteria ATSP (bi-ATSP) which is a special case
of the multicriteria ATSP~\cite{Ehrgott}, where an arc is associated to a couple
of weights.

The best possible solution to a multicriteria optimization problem (MOP) is usually supposed
to be the Pareto set~\cite{Ehrgott,Podinovskiy_Noghin},
which is rather wide in real-world problems, and difficulties arise in choosing the final variant.
For that reason numerous methods introduce some mechanism to treat the MOP:
utility function, rule, or binary relation, so that methods are aimed at finding
an ``optimal'' solution with respect to this mechanism.
However, some approaches do not guarantee that the obtained solution will be from the Pareto set.
State-of-the-art methods are the following~\cite{Figueira_Greco_Ehrgott}:
multiattribute utility theory, outranking approaches, verbal decision analysis,
various iterative procedures with man-machine interface, etc.
In this paper, we investigate the axiomatic approach of the Pareto set reduction proposed in~\cite{Noghin2018} which has an alternative idea.
Here the author introduced an additional information about the decision maker (DM) preferences in terms of
the so-called ``quantum of information''.
The method shows how to construct a new bound of the optimal choice, which is narrower than the Pareto set.
Practical applications of the approach could be found in~\cite{Klimova,Noghin_Prasolov}.

As far as we know, the axiomatic approach of the Pareto set reduction has not been widely investigated
in the case of discrete optimization problems,
and an experimental evaluation has not been carried out on real-world instances.
We apply this approach to the bi-ATSP in order to estimate its effectiveness,
i.e. the degree of the Pareto set reduction and how it depends on the parameters of the information about DM's preferences.
We identify series of ``quanta of information'' that guarantee the reduction of the Pareto set
for particular cases of the bi-ATSP.
\looseness=-1

Originally the reduction is constructed with respect the Pareto set of the considered problem.
Due to the strongly NP-hardness of the bi-ATSP we take an approximation of the Pareto set in computational experiments.
The ATSP cannot be approximated
with any constant or exponential approximation factor already with a single objective function~\cite{ACGKMP2005}.
Moreover, in~\cite{ABGM2005}, the non-approximability bounds were obtained for the multicriteria ATSP with weights 1 and 2.
The results are based on the non-existence of a small size approximating set.
Therefore, metaheuristics, in particular multi-objective evolutionary algorithms (MOEAs),
are appropriate to approximate the Pareto set of the bi-ATSP.

Numerous MOEAs have been proposed to MOPs
(see e.g. \cite{Buzdalov2015,Deb,LiZhang2009,YXW2014,ZBT2007,ZLT2001}).
There are three main classes of approaches to develop MOEAs, which are known as
Pareto-dominance based (see e.g. SPEA2~\cite{ZLT2001},  NSGA-II~\cite{Buzdalov2015,Deb}, NSGA-III~\cite{YXW2014}),
decomposition based (see e.g. MOEA/D~\cite{LiZhang2009}) and indicator based approaches (see e.g. SIBEA~\cite{ZBT2007}).
NSGA-II~\cite{Deb} has one of the best results in the literature on
multi-objective genetic algorithms (MOGAs) for the MOPs with two or three
objectives.
In~\cite{Buzdalov2015}, a fast implementation of a steady-state version of NSGA-II is proposed for two dimensions.

In~\cite{GCH2007,PDM2015}, NSGA-II was adopted to the multicriteria symmetric traveling salesman problem, and
the experimental evaluation was performed on symmetric instances from TSPLIB library~\cite{Reinelt91}.
To the best of our knowledge, there is no adaptation of NSGA-II to the more general
problem, where arc weights are non-symmetric.
In this paper, we propose a new MOGA based on NSGA-II to solve the bi-ATSP using
adjacency-based representation of solutions.
A computational experiment is carried out on randomly generated instances.
The results of the experiment show the degree of the reduction of the Pareto set approximation
for various ``quanta of information'' and various structures of the problem instances.
\looseness=-1

\section{Problem Statement} \label{sec:Problem}

An instance of the traveling salesman problem~\cite{ACGKMP2005} (TSP) is given
by a complete graph $G = (V,E)$, where $V=\{v_1,\dots,v_n\}$ is the set of vertices
and set $E$ contains arcs (or edges) between every pair of vertices in $V$.
Each arc (or edge) $e\in E$ is associated  with a weight $d(e)$.
The aim is to find a Hamiltonian circuit  (also called a tour) of minimum weight,
where the weight of a tour $C$ is the sum of its arc (or edge) weights $\sum_{e\in C}d(e)$.
We denote by $\mathcal{C}$ all possible $(n-1)!$ tours of graph $G$.
If graph $G$ is undirected, we have Symmetric TSP (STSP).
If $G$ is a directed graph, then we have Asymmetric TSP (ATSP).

In many situations, however, there is more than one objective function (criterion) to optimize~\cite{Ehrgott,Podinovskiy_Noghin}.
In case of the TSP, we might want to minimize the travel distance, the travel time, the expenses,
the number of flight changes, etc.
This gives rise to a multicriteria TSP, where Hamiltonian circuits are sought that optimize several objectives simultaneously.
For the $m$-criteria TSP, each arc (or edge) $e$ has a weight $d(e)= (d_1(e),\dots,d_m(e))$,
which is a vector of length $m$ (instead of a scalar).
The total weight of a tour $C$ is also a vector $D(C)=(D_1(C),\dots,D_m(C))$,
where $D_j(C) =\sum_{e\in C} d_j(e),\ j=1,\dots,m$.
Given this, the goal of the optimization problem could be the following:
find a feasible solution which simultaneously minimizes each coordinate.
Unfortunately, such an ideal solution rarely exists since objective functions are normally in conflict.

We say that one solution (tour) $C^\ast$ dominates another solution $C$
if the inequality $D(C^\ast) \leq D(C)$ holds. The notation $D(C^\ast) \leq D(C)$ means
that $D(C^\ast) \neq D(C)$ and $D_i(C^\ast) \leqslant D_i(C)$
for all $i \in I$, where $I = \{1, 2, \ldots, m \}$. This relation $\leq$ is also called {\it the Pareto relation}.
A set of non-dominated solutions is called {\it  the set of pareto-optimal solutions}~\cite{Ehrgott,Podinovskiy_Noghin}
$P_D(\mathcal{C})=\{ C \in \mathcal{C} \mid \nexists C^\ast \in \mathcal{C}: D(C^\ast)~\leq~D(C)~\}.$
In discrete problems, the set of pareto-optimal solutions is non-empty if the set of feasible solutions is non-empty,
which is true for the multicriteria TSP.
If we denote $\mathcal{D} = D(\mathcal{C})$, then {\it the Pareto set} is defined as
$P(\mathcal{D})=\{ y \in \mathcal{D} \mid \nexists y^\ast \in \mathcal{D}: y^\ast~\leq~y~\}.$
We assume that the Pareto set is specified except for a collection of equivalence classes,
generated by equivalence relation $C' \sim C''$ iff $D(C') = D(C'')$.

In this paper, we investigate the issue of the Pareto set reduction for the bi-ATSP.

\section{Pareto Set Reduction} \label{sec:PS_Reduction}

Axiomatic approach of the Pareto set reduction is applied to both discrete and continuous problems.
Due to consideration of the multicriteria ATSP we formulate the basic concepts and results of the approach in terms of notations
introduced in Section~\ref{sec:Problem}.
Further, we investigate properties of the bi-ATSP in the scope of the Pareto set reduction.

\subsection{Main Approach} \label{subsec:PS_Reduction_MA}

According to~\cite{Noghin2018} we consider the extended multicriteria problem $<\mathcal{C}, D, \prec>$:
\begin{itemize}
\item a set of all possible $(n-1)!$ tours $\mathcal{C}$;
\item a vector criterion $D = (D_1, D_2, \ldots,  D_m)$ defined on set $\mathcal{C}$;
\item an asymmetric binary preference relation of the DM $\prec$ defined on set $\mathcal{D}$.
\end{itemize}
The notation $D(C') \prec D(C'')$ means that the DM prefers the solution $C'$ to $C''$.

Binary relation $\prec$ satisfies some axioms of the so-called ``reasonable'' \ choice, according which
it is irreflexive, transitive,
invariant with respect to a linear positive transformation and compatible with each criteria $D_1, D_2, \ldots, D_m$.
The compatibility means that the DM is interested in
decreasing value of each criterion when values of other criteria are constant.
Also, if for some
feasible solutions $C', \ C'' \in \mathcal{C}$
the relation $D(C') \prec D(C'')$ holds, then tour $C''$
does not belong to the optimal choice within the whole set $\mathcal{C}$.

In~\cite{Noghin2018}, the author established the Edgeworth--Pareto principle:
under axioms of ``reasonable'' choice any set of selected outcomes $Ch(\mathcal{D})$
belongs to the Pareto set $P(\mathcal{D})$.
Here the set of selected outcomes is interpreted as some abstract set
corresponded to the set of tours, that satisfy all hypothetic preferences of the DM.
So, the optimal choice should be done within the Pareto set only
if preference relation $\prec$ fulfills the axioms of ``reasonable'' choice.

In real-life multicriteria problems the Pareto set is rather wide.
For this reason V.~Noghin proposed a specific information on the DM's preference relation $\prec$ to reduce
the Pareto set staying within the set of selected outcomes~\cite{Noghin2014,Noghin2018}:

\begin{definition}
\label{ref_def_i_j}
We say that there exists a {\it ``quantum of information''} about the DM's preference
relation $\prec$ if vector $y' \in \R^m$ such that
$y'_i = - w_i < 0$, $y'_j = w_j > 0$,
$y'_s = 0$ for all $s \in I\setminus \{i, j\}$
satisfies the expression $y' \prec 0_m$.
In such case we will say, that the component of criteria $i$ is more important than the component~$j$
with given positive parameters~$w_i$,~$w_j$.
\end{definition}

Thus, ``quantum of information'' shows that the DM is ready to compromise
by increasing the criterion $D_j$ by amount $w_j$ for decreasing the criterion $D_i$ by amount $w_i$.
The quantity of relative loss is set by the so-called
coefficient of relative importance $\theta = w_j / (w_i + w_j)$, therefore $\theta \in (0, \ 1)$.

As mentioned before the relation $\prec$ is invariant with respect to a linear positive transformation.
Hence Definition~\ref{ref_def_i_j} is equivalent to the existence of such vector $y'' \in \R^m$
with components $y''_i = \theta - 1$, $y''_j = \theta$, $y''_s = 0$ for all $s \in I\setminus \{i, j\}$
that the relation $y'' \prec 0_m$ holds. Further, in experimental study (Section~\ref{sec:experiments})
we consider ``quantum of information'' exactly in terms of coefficient $\theta$.

In~\cite{Noghin2018}, the author established the rule of taking into account ``quantum of information''.
This rule consists in constructing a ``new'' vector criterion using the components of the ``old'' one
and parameters of the information $w_i$, $w_j$. Then one should find the Pareto set of ``new'' multicriteria problem
with the same set of feasible solutions and ``new'' vector criterion.
The obtained set will belong to the Pareto set of the initial problem
and give a narrower upper bound on the optimal choice, as a result the Pareto set will be reduced.

The following theorem states the rule of applying ``quantum of information'' and
specifies how to evaluate ``new'' vector criterion upon the ``old'' one.

\begin{theorem}[\cite{Noghin2018}]
\label{th_red}
Given a ``quantum of information'' by Definition~\ref{ref_def_i_j},
the inclusions $Ch(\mathcal{D}) \subseteq \hat{P}(\mathcal{D}) \subseteq P(\mathcal{D})$ are valid
for any set of selected outcomes $Ch(\mathcal{D})$.
Here $\hat{P}(\mathcal{D}) = D(P_{\hat{D}}(\mathcal{C}))$, and $P_{\hat{D}}(\mathcal{C})$ is the set of pareto-optimal solutions
with respect to $m$-dimensional vector criterion $\hat{D} = (\hat{D}_1, \ldots, \hat{D}_m)$, where
$\hat{D}_j = \theta D_i + (1-\theta) D_j$, $\hat{D}_s = D_s$ for all $s \neq j$.
\end{theorem}

Thus, ``new'' vector criterion $\hat{D}$ differs from the ``old'' one
only by less important component $j$.
In~\cite{Klimova_Noghin2006,Noghin2011,Zakharov2012} one can find results
on applying particular collections of ``quanta of information'' and
scheme to arbitrary collection.

\subsection{Pareto Set Reduction in Bi-ATSP}
\label{subsec:PS_Reduction_prop}

Here we consider the bi-ATSP and its properties with respect to reduction of the Pareto set.

Obviously, the upper bound on the cardinality of the Pareto set $P(\mathcal{D})$ is $(n-1)!$,
and this bound is tight~\cite{Emelichev_Perepeliza}.
In~\cite{Vinogradskaya_Gaft} authors established the maximum number of elements in the Pareto set
for any multicriteria discrete problem,
that in the case of the bi-ATSP gives the following upper bound:
$|P(\mathcal{\mathcal{D}})| \leqslant \min\{l_1, l_2\}$, where $l_i$ is the number of different values in
the set $\mathcal{D}_i = D_i(\mathcal{C})$, $i = 1, 2$.
In the case of the bi-ATSP with integer weights we get
$l_i \leqslant \max\{\mathcal{D}_i\} - \min\{\mathcal{D}_i\} + 1$,
where values $\max\{\mathcal{D}_i\}$ and $\min\{\mathcal{D}_i\}$ can be replaced
by upper and lower bounds on the objective function ${D}_i$, $i = 1, 2$. 

Now, we go to establish theoretical results estimating the degree of the Pareto set reduction.
Let us consider the case when all elements of the Pareto set lay on principal diagonal of some rectangle in the criterion space.

\begin{theorem}
\label{prop1_crit_theta}
Let $P(\mathcal{D}) = \{ (y_1, y_2): y_2 = a - k y_1, y_1 \in \mathcal{D}_1, y_2 \in \mathcal{D}_2 \}$,
where $a$ and $k$ are arbitrary positive constants.
Suppose the 1st criterion $D_1$ is more important than the 2nd one $D_2$
with coefficient of relative importance~$\theta'$. If $\theta' \geqslant k/(k+1)$, then
the reduction of the Pareto set $\hat{P}(\mathcal{D})$ consists of only one element.
In the case of $\theta' < k/(k+1)$ the reduction does not hold, i.e. $\hat{P}(\mathcal{D}) = P(\mathcal{D})$.
\end{theorem}

\begin{theorem}
\label{prop2_crit_theta}
Let in Theorem~\ref{prop1_crit_theta}, otherwise,
the 2nd criterion $D_2$ is more important than the 1st one $D_1$
with coefficient of relative importance $\theta''$.
Then the reduction of the Pareto set $\hat{P}(\mathcal{D})$ has only one element if $\theta'' \geqslant 1/(k+1)$,
and $\hat{P}(\mathcal{D}) = P(\mathcal{D})$ if $\theta'' < 1/(k+1)$.
\end{theorem}

Particularly, if the feasible set $\mathcal{D}$ lay on the line $y_2 = a - k y_1$,
we have $P(\mathcal{D}) = \mathcal{D}$,
and the conditions of Theorems~\ref{prop1_crit_theta} and~\ref{prop2_crit_theta} hold.
In such case we say, that {\it criteria $D_1$ and $D_2$ contradict each other with coefficient~$k$}.

Obviously, for any bi-ATSP instance there exists the minimum number of parallel lines with a negative slope,
that all elements of the Pareto set belong to them. Thus we have

\begin{corollary}
\label{cl_crit_theta}
Let $P(\mathcal{D}) = \bigcup_{i=1}^p \{ (y_1, y_2): y_2 = a_i - k y_1, y_1 \in \mathcal{D}_1, y_2 \in \mathcal{D}_2 \}$,
where $a_i$, $i = 1, \ldots, p$, and $k$ are arbitrary positive constants.
If criterion $D_1$ is more important than criterion $D_2$
with coefficient of relative importance~$\theta'$ and $\theta' \geqslant k/(k+1)$,
or criterion $D_2$ is more important than criterion $D_1$
with coefficient of relative importance~$\theta''$ and $\theta'' \geqslant 1/(k+1)$,
then $|\hat{P}(\mathcal{D})| \leqslant p$.
\end{corollary}

Further, we identify the condition that guarantees excluding at least one element from the Pareto set.
\begin{proposition}
\label{prop_one_point}
Let the criterion $D_i$ is more important than the criterion $D_j$
with coefficient of relative importance $\theta$.
Suppose that there exist such tours $C', C'' \in P_D(\mathcal{C})$ that the following inequality holds:
\begin{equation}
\label{prop_2}
\frac{D_i(C') - D_i(C'')} {D_j(C'') - D_j(C')} \geqslant \frac{1 - \theta} {\theta},
\end{equation}
then $|P(\mathcal{D})| - |\hat{P}(\mathcal{D})| \geqslant 1$. 
Here $i, j \in \{1, 2\}$, $i \neq j$.
\end{proposition}

The difficulty in checking inequality~(\ref{prop_2}) is that we should know two elements of the Pareto set.
Meanwhile the tours $C_{min_1} = \mbox{argmin}\{D_1(C), \ {C \in \mathcal{C}}\}$,
$C_{min_2} = \mbox{argmin}\{D_2(C), \ {C \in \mathcal{C}}\}$ are pareto-optimal by definition.

The proofs of Theorems \ref{prop1_crit_theta}, \ref{prop2_crit_theta} and Proposition~\ref{prop_one_point}
are based on geometrical representation of the Pareto set reduction~\cite{Noghin2018}.
The results of this subsection are true for any discrete bicriteria problem.

\section{Multi-Objective Genetic Algorithm} \label{sec:MOGA}

The genetic algorithm is a random search method
that models a process of evolution of a population of {\em
individuals}~\cite{Reev1997}. Each individual is a sample solution
to the optimization problem being solved.
Individuals of a new population are built by means of reproduction
operators (crossover and/or mutation).

\subsection{NSGA-II Scheme}\label{subsec:NSGA-II}

To construct an approximation of the Pareto set
to the bi-ATSP we develop a MOGA based on Non-dominated Sorting Genetic Algorithm II (NSGA-II)~\cite{Deb}.
The NSGA-II is initiated by generating $N$ random solutions of the initial population.
Then the population is sorted based on the non-domination relation (the Pareto relation).
All individuals of the population which
are not dominated by any other individual compose the first {\em non-dominated level} and are marked with the {\em rank} of 1,
all individuals which are dominated by at least one individual of the rank $i-1$ compose the $i$-th non-dominated level and
are marked with the rank of $i$, $i=2,\ 3,\ \dots$.
To get an estimate of the density of solutions surrounding a solution $x$ in a non-dominated level of the population,
two nearest solutions on each side of this solution are identified for each of the objectives.
The estimation of solution $x$ is called {\em crowding distance} and 
it is computed as a normalized perimeter of the cuboid formed in the criterion space by
the nearest neighbors.
\looseness=-1

The NSGA-II is characterized by the population management strategy known as generational model~\cite{Reev1997}.
Here the next population $P_{t}$ is constructed from the best $N$ solutions of the current population~$P_{t-1}$ and
an offspring population $Q_{t-1}$ created from $P_{t-1}$ by applying selection, crossover, and mutation. %
The best solutions are selected using the rank and the crowding distance.
Between two solutions with differing non-domination ranks, we prefer the solution with the lower rank.
If both solutions belong to the same level, then we prefer the solution with the bigger crowding distance. 
The formal scheme of the NSGA-II is as follows:
\looseness=-1

{\bf Non-dominated Sorting Genetic Algorithm II}

\parshape=3 0em 35em 1.5em 33.5em 1.5em 33.5em
{\sc Step~1}.~Construct the initial population $P_0$ of size $N$ and assign $t:=1$. The
population~$P_0$ is sorted based on the non-domination relation. The crowding distan\-ces
of individuals are calculated.

{\sc Step~2}.~Repeat steps~2.1-2.4 until some stopping criterion is satisfied:

\mbox{\hspace{1.5em}} {\sc 2.1.}~Create offspring population $Q_{t-1}$.

\mbox{\hspace{1.5em}} Steps 2.1.1-2.1.4 are performed $N$ times:

\mbox{\hspace{2.5em}} {\sc 2.1.1.}~Choose two parent individuals~${\bf p}_1, {\bf p}_2$
from the population.

\mbox{\hspace{2.5em}} {\sc 2.1.2.}~Apply mutation to ${\bf p}_1$ and ${\bf p}_2$ and obtain
individuals~${\bf p}'_1, {\bf p}'_2$.

\mbox{\hspace{2.5em}} {\sc 2.1.3.}~Create an offspring ${\bf p}'$, applying a crossover
to~${\bf p}'_1$ and ${\bf p}'_2$.

\mbox{\hspace{2.5em}} {\sc 2.1.4.}~Put individual ${\bf p}'$ into population $Q_{t-1}$.

\parshape=3 1.85em 33.15em 3.5em 31.5em 3.5em 31.5em
{\sc 2.2.}~Form a combined population $R_{t-1}:=P_{t-1}\cup Q_{t-1}$.
The population~$R_{t-1}$ is sorted based on the non-domination relation. The crowding distances of individuals are calculated.

\parshape=2 1.85em 33.15em 3.5em 31.5em
{\sc 2.3.}~Construct population $P_t$ from the best individuals of population
$R_{t-1}$ using the rank and the crowding distance to select solutions.

\mbox{\hspace{1.5em}} {\sc 2.4.}~Set $t:=t+1$.

One iteration of the presented NSGA-II is performed in $O(mN^2)$ time as shown in~\cite{Deb}.
In our implementation of the NSGA-II
four individuals of the initial population are constructed by a problem-specific heuristic presented in~\cite{EK2018}
for the ATSP with one criterion. The heuristic first solves
the Assignment Problem, and then patches the circuits of the optimum
assignment together to form a feasible tour in two ways.
So, we create two solutions with each of the objectives.
All other individuals of the initial population are generated randomly.

Each parent on Step~2.1.1 is chosen by $s$-{\em tournament
selection}: sample randomly $s$~individuals from the current
population and select the best one by means of the rank and the crowding distance.

\subsection{Recombination and Mutation Operators}\label{subsec:recombinationandmutation}

The experimental results of~\cite{EK2018,S91} for the TSP indicate
that  reproduction operators with the adjacency-based representation of solutions have an
advantage over operators, which emphasize the order or position of the vertices in parent solutions.
We suppose that a feasible solution to the bi-ATSP  is encoded as
a list of arcs.
In the recombination operator on Step 2.1.3 we use a variant of the {\em Directed Edge Crossover} (DEC), 
which may be considered as a  ``direct descendant'' of  Edge Crossover~\cite{S91}
originally developed for the STSP.
\looseness=-1

The DEC operator is {\em respectful}~\cite{R94},
i.e. all arcs shared by both parents are copied into the offspring.
The remaining arcs are selected so as the  preference is given to those arcs
that are contained in at least one of the parents.
Arcs are inserted taking into account the non-violation of sub-tour elimination constraints.
If the obtained offspring is equal to one of the parents,
then the result of the recombination is calculated by applying the well-known {\em shift mutation}~\cite{R94}
to one of the two parents with equal probability.
This approach allows us to avoid creating a clone of parents and to maintain a diverse set of solutions in the population.
\looseness=-1

The mutation is also applied to each parent on Step 2.1.2 with
probability $p_{mut}$, which is a tunable parameter of the MOGA.
We use a mutation operator proposed in~\cite{EK2018} for the one-criteria ATSP.
It performs a random jump within 3-opt neighborhood,
trying to improve a parent solution in terms of one of the criteria.
Each time one of two objectives is used in mutation with equal probability.
\looseness=-1

\section{Computational Experiment} \label{sec:experiments}

This section presents the results of the computational experiment on the bi-ATSP instances.
Our MOGA (NSGA-II-biATSP) was programmed in C++ and tested on a
computer with Intel~Core~i5~3470 3.20~GHz processor, 4 Gb~RAM.

Various meta-heuristics and heuristics have been developed for the multicriteria STSP,
such as Pareto local search algorithms, MOEAs, multi-objective ant colony optimization methods,
memetic algorithms and others (see, e.g., \cite{GCH2007,JZ2009,KS2007,LT2010,PDM2015}).
However, we have not found in the literature any multi-objective metaheuristic proposed specifically
to the  multicriteria ATSP and experimentally tested on instances with non-symmetric weights of arcs.

We carried out the preliminary study to evaluate the performance
of our GA on bi-ATSP instances generated randomly with $n=12$.
The Pareto sets were found by an exact algorithm~\cite{Noghin2018}.
The generational distance~\cite{YXW2014} and
the inverted generational distance~\cite{YXW2014} were involved as performance metrics.
The experimental evaluation showed that the proposed MOGA  yields competitive results.
The values of metrics decrease not less than 7 times during $5000$ iterations,
and the final values are approximately $0.6$ on average.
The number of elements in the final approximation is at least $80\%$ of $|P(\mathcal{D})|$.
This indicates the convergence of the approximation obtained by NSGA-II-biATSP to the Pareto set and its diversity.
Here the detailed description of  the preliminary study is omitted,
as the main goal of the paper is to investigate the axiomatic approach of the Pareto set reduction in the case of bi-ATSP.
\looseness=-1

Note that there exists MOOLIBRARY library~\cite{moolib},
which contains  instances of some discrete multicriteria problems. 
However, the multicriteria TSP is not presented in this library,
so we generate the bi-ATSP test instances randomly and construct them from the ATSP instances of TSPLIB library, as well.

The reduction of the Pareto set approximation was tested on
the following medium-size problem instances of four series with {$n=50$}: S50[1,10][1,10], S50[1,20][1,20], S50[1,10][1,20], S50contr[1,2][1,2].
Each series consists of five problems with integer weights $d_1(\cdot)$ and $d_2(\cdot)$
of arcs randomly generated from intervals specified at the ending of the series name.
In series S50contr[1,2][1,2] the criteria contradict each other with coefficient~$1$, i.e. weights are generated so that $d_2(e)=3-d_1(e)$ for all $e\in E$.
We also took seven ATSP instances of series ftv from TSPLIB library~\cite{Reinelt91}:
ftv33, ftv35, ftv38, ftv44, ftv47, ftv55, ftv64.
The ftv collection includes instances from vehicle routing applications~\cite{Reinelt91}.
These instances compose series denoted by SftvRand, and their arc weights are used for the first criterion.
The arc weights for the second criterion are generated randomly from interval $[1,d_1^{\max}]$,
were $d_1^{\max}$ is the maximum arc weight on the first criterion.
We set the population size $N=100$,  the tournament size~${s=10}$, and the mutation probability~${p_{mut}=0.1}$.
To construct an approximation of the Pareto set $A$ for each instance we run NSGA-II-bi-ATSP once
and the run continued for $5000$ iterations.

We compare two cases when the 1st criterion is more important than the 2nd criterion (1st-2nd case),
and vice versa (2nd-1st case).
The degree of the reduction of the Pareto set approximation was investigated
with respect to coefficient of relative importance varying from $0.1$ to $0.9$ by step $0.1$.
On all instances
for each value of $\theta$ we re-evaluate the obtained approximation in terms of ``new'' vector criterion $\hat{D}$
upon the formulae from Theorem~\ref{th_red}. Then by the complete enumeration we find the Pareto set approximation
in ``new'' criterion space that gives us the reduction of the Pareto set approximation in the initial criterion space.
\looseness=-1

The number $N^A$ of elements of the Pareto set approximation $A$ and
the percentage of the excluded elements from set $A$
are presented on average over series in Tables~\ref{tab:1st2nd} and \ref{tab:2nd1st}.
Let $\Delta_i$ be the difference between the maximum and minimum values of
the Pareto set approximation on the $i$-th criterion, $i = 1, 2$.
The value $\delta_{21} = \Delta_2 / \Delta_1$ indicates the ratio between diversities of criteria of set~$A$.
\looseness=-1

\begin{table}[!h]
\caption{Reduction of the Pareto set approximation  in the 1st-2nd case}
\label{tab:1st2nd}
\begin{tabular}{|c|c|c|c|c|c|c|c|c|c|c|c|}
\hline
Series & \multicolumn{9}{c|}{$\theta$} & $N^A$ & $\delta_{21}$ \\
\cline{2-10}
&0.1&0.2&0.3&0.4&{\bf 0.5}&0.6&0.7&0.8&0.9&aver& aver\\
\hline
S50contr[1,2][1,2] & 0& 0& 0& 0& {\bf 98.04} &  98.04  & 98.04  & 98.04 &  98.04 & 51 & 1 \\
\hline
S50[1,10][1,10] & 4.42 &   17.76 &  41.64 &  60.43 &  {\bf 72.32} &  78.61 &  90.29  & 95.92  & 97.78 & 45.8 &1.02 \\
\hline
S50[1,20][1,20] & 5.97 &  23.09 &  38.15 &  59.92 &  {\bf 73.69} &  79.91 &  90.18 &  94.71 &  98.05 & 57.4 & 1.07 \\
\hline
SftvRand & 6.63 & 16.47 & 27.99 &  41.31 & {\bf 58.31}  & 72.96 & 86.24 & 93.7 & 96.71  & 61.86  & 1.56 \\
\hline
S50[1,10][1,20] & 2.19  & 9.02 &   19.18 &  28.3 &  {\bf 45.91} &  61.69 &  71.79 &  83.39 &  95.46 & 51.6 & 2.08 \\
\hline
\end{tabular}
\end{table}

\begin{table}[!h]
\caption{Reduction of the Pareto set approximation in the 2nd-1st case}
\label{tab:2nd1st}
\begin{tabular}{|c|c|c|c|c|c|c|c|c|c|c|c|}
\hline
Series & \multicolumn{9}{c|}{$\theta$} & $N^A$ & $\delta_{21}$ \\
\cline{2-10}
&0.1&0.2&0.3&0.4&{\bf 0.5}&0.6&0.7&0.8&0.9&aver& aver\\
\hline
S50contr[1,2][1,2] & 0& 0& 0& 0& {\bf 98.04} &  98.04  & 98.04  & 98.04 &  98.04 & 51 & 1 \\
\hline
S50[1,10][1,10] & 3.7 & 19.8 & 37.92 &  52.85 &  {\bf 67.38}  & 80.58 &  92.92 & 97.32 &  97.32 & 45.8 &1.02\\
\hline
S50[1,20][1,20] & 7.79  & 21.1 &  36.73  & 54.5 &  {\bf 69.8} &  82.93 &  93.34 & 97.2 &   97.91 & 57.4 & 1.07 \\
\hline
SftvRand & 17.42 & 36.73 & 59 & 74.33 & {\bf 87.37} & 92.04 & 97.67 & 98.11 & 98.36 & 61.86  & 1.56 \\
\hline
S50[1,10][1,20] & 19.91 &  42.97 &  62.21 &  77.39 &  {\bf 92.41} &  95.54 & 97.15 &  98.02 &  98.02 & 51.6 & 2.08 \\
\hline
\end{tabular}
\end{table}

For series S50[1,10][1,10], S50[1,20][1,20]
when $\theta=0.5$ approximately $70 \%$ of elements of the set $A$ are excluded,
and when $\theta=0.7$ less than $10 \%$ of elements are remained.
The statement is valid for both 1st-2nd and 2nd-1st cases.
Series SftvRand shows different results: in the 1st-2nd case the reduction occurs ``almost uniformly'', i.e.
the value of $\theta$ is almost proportional to the degree of the reduction,
in the 2nd-1st case the condition $\theta = 0.5$ gives approximately $90 \%$ of the excluded elements.
On series S50[1,10][1,20] in the 1st-2nd case the degree of the reduction grows slowly as $\theta$ tends to 1
in comparison to other series, and in the 2nd-1st case more than $90 \%$ of elements are eliminated at $\theta = 0.5$.

Also, we note that on series S50[1,10][1,20] (SftvRand) for $\theta = 0.5$ the percentage of the excluded elements
in the 2nd-1st case is approximately $2$ ($1.5$) times
greater than the percentage of the excluded elements in the 1st-2nd case.
Note that  $\delta_{21} \approx 2$ for series S50[1,10][1,20] and
$\delta_{21} \approx 1.5$ for series SftvRand.
Therefore, the ratio between diversities of values of the Pareto set approximation on components of criteria
influences on the degree of the reduction in the same proportion when $\theta = 0.5$
(each criterion has relatively the same importance).

On series S50contr[1,2][1,2], where the components of criterion contradict each other with coefficient~$1$,
we do not have a reduction when $\theta < 0.5$, and the reduction up to one element takes place when $\theta \geqslant 0.5$.
Thus, the results of the experiment confirm the theoretical results of Subsection~\ref{subsec:PS_Reduction_prop}.
Moreover, identical character of the reduction for both 1st-2nd and 2nd-1st cases occurs only
on series S50[1,10][1,10], S50[1,20][1,20], and S50contr[1,2][1,2], which have the same diversity and distribution
with respect to both criteria.

Based on the results of the experiment we suppose that
the degree of the reduction of the Pareto set approximation will be similar
for the large-size problems with the same structure as the considered instances.

\section{Conclusion}\label{sec:concl}

We applied to the bicriteria ATSP  the axiomatic approach of the Pareto set reduction proposed by V.~Noghin.
For particular cases the series of ``quanta of information''
that guarantee the reduction of the Pareto set were identified.
An approximation of the Pareto set to the bicriteria ATSP was found by a new generational multi-objective genetic algorithm.
The experimental evaluation indicated the degree of reduction of the Pareto set approximation for various ``quanta of information'' and various problem structures.

Further research may include construction and analysis of new classes of multicriteria ATSP instances
with complex structures of the Pareto set.
It is also important to consider real-life ATSP instances with real-life decision maker
and investigate effectiveness of the axiomatic approach for them.
Moreover, developing a faster implementation of the multi-objective genetic algorithm
with steady-state replacement and local search procedures has great interest.\\

\noindent
\textbf{\ackname} The research was supported by RFBR grant 17-07-00371 (A.~Zakharov)
and by the Ministry of Science and Education of the Russian Federation under the 5-100 Excellence Programme (Yu.~Kovalenko).


\end{document}